\documentclass[final,12pt]{elsarticle}

\usepackage{graphicx}
\usepackage{amsmath} 
\usepackage{amssymb}
\biboptions{numbers,sort&compress}
\usepackage{siunitx}
\usepackage{color} \usepackage{cancel}
\usepackage{booktabs,caption}

\def\figref#1{Fig.~\ref{#1}}
\def\tabref#1{Tab.~\ref{#1}}
\def\eqnref#1{Eq.~(\ref{#1})}

\journal{Journal of Luminescence} 
\begin{document}

\begin{frontmatter}

\title{{\bf Infrared and visible scintillation of
 Ho$^{3+}$-doped YAG and YLF crystals}}
 
\author[mymainaddress]{F. Chiossi\corref{mycorrespondingauthor}}
\address[mymainaddress]{CNISM Unit, Department of Physics and Astronomy, University of Padua and\\
Istituto Nazionale Fisica Nucleare, sez. Padova\\ via F-Marzolo 8, I-35131 Padua, Italy}
\cortext[mycorrespondingauthor]{Corresponding author}
\ead{federico.chiossi@phd.unipd.it} 
 
\author[mymainaddress]{A. F. Borghesani}

\author[mysecondaryaddress]{G. Carugno}
\address[mysecondaryaddress]{Department of Physics and Astronomy, University of Padua and \\
Istituto Nazionale Fisica Nucleare, sez. Padova \\via F-Marzolo 8, I-35131 Padua, Italy}

\begin{abstract}
In our effort to develop a new kind of detector for low-energy, low-rate energy deposition events we have investigated the cathodo- and radioluminescence of Ho:YAG and Ho:YLF single crystals in an extended wavelength range from $200\,$nm to $2200\,$nm. The emission spectra of both crystals show a much more intense emission in the infrared range than in the visible one. We estimate an infrared light yield
of $40\,$photons/keV when exciting the crystals with X-rays of energy $\approx 30\,$keV. The main reason of this high value is due to the Ho$^{3+}$ ions energy levels scheme that allows efficient cross relaxation processes to occur even at low dopant concentration.
 
\end{abstract}


\end{frontmatter}

\section{Introduction}\label{sec:intro}

Inorganic scintillators with very high light yield (LY) have been discovered in the past two decades, such as Ce:LuI$_3$ and Eu:SrI$_2$.
However, the energetic efficiency of the electron-hole pair creation sets a fundamental upper limit on the LY.
This is equal to $\approx 10^6/(\beta E_g)$\,photons/MeV, where $E_g$ is the scintillator band gap and $2\le\beta\le 3$ is a constant~\cite{dorenbosintro}. In order to overcome this limit, we have started investigating the feasibility of  laser induced scintillation enhancement in rare earth (RE) doped crystals.

In the particular combination of excited state laser absorption and cross relaxation, known as photon avalanche upconversion 
\cite{avalanche, avalanche2}, the emission of several photons can be accomplished by starting from one single excited RE ion.
In the cross relaxation process, a high energy excited RE ion shares fraction of its energy to another one in the ground state. In this way, it is even possible to obtain two RE ions in the same state. A suitably tuned laser can promote these two excited ions to the former high energy state, in which cross relaxation is likely to occur once more, thereby accomplishing an avalanche process. 

This scheme has been implemented for the realization of a low noise infrared radiation detector~\cite{Pr_detector}. We are investigating if
it is applicable also for the rare- and low energy- deposition event detection, such as dark matter and nucleus-neutrino scattering research, in which a fast scintillator response is not necessary. We have already proved that the laser induced upconversion process can shift scintillation to a shorter wavelength, which can be detected with  higher quantum efficiency and with lower dark count rate than in the infrared range~\cite{Borghesani2015}. 

However, at this point of our research, the intrinsic noise of this kind of detector is still too high for any practical applications~\cite{Borghesani2015,Braggioaxion}. 
Although the laser wavelength does not correspond to any pure electronic transitions from the ground state, the laser light can still be 
absorbed by means of several mechanisms, leading to populate the same level that
we would like to be populated only by particle excitation.

In order to reduce this laser induced noise background, we are looking for different RE dopant, matrix and pump scheme combinations. At the same time we are also attempting at proving that low energy RE manifolds are efficiently populated by energetic radiation. To this purpose, we use electron impact- and X-rays excitations to investigate the visible and infrared scintillation of different RE-matrix combination. We have already reported a relatively high LY yield for Nd:YAG (1.1\%) and Tm:YAG (4.4\%)\,\cite{TmNd} and in this paper we present
the results for Holmium doped YAG and YLF crystals.

\section{Experimental Apparatus and Method}\label{sect:app}

Experimental apparatus and procedures have previously been described in detail~\cite{Pr,TmNd}. Here, we only recall their principal features. 

A home-made electron gun is able to deliver a current up to $15\,\mu$A of electrons with energy of 70\,keV~\cite{borghesani2011}. In the pulsed mode, electron bunches of several nC as short as $100\,\mu$s can be released with an adjustable repetition rate between 20 and 1000\,Hz. A thin Titanium, or Tantalum foil, is placed in
front of the sample and allows the recollection and analysis of the electrons accelerated towards the crystals. 

 In the Titanium case, the electrons 
cross the foil with an energy loss estimated to be $\approx$15\,keV.
In the Tantalum case, the electrons are stopped in the foil and X-rays are produced. A Carbon foil, placed between the metal foil and the crystals, filters out the lower energy X-rays component.  

We have studied a home-grown Ho$^{3+}$:YLiF$_4$ 0.8\%\,at (7x7x8\,mm$^3$) and a commercial Ho$^{3+}$:Y$_3$Al$_5$O$_{12}$ 2.5\%\,at (3\,mm in height, 5\,mm in diameter) single crystals.

The crystals cathodo- and radioluminescence are detected with Silicon (Ha\-ma\-mat\-su mod. S1337-1010BQ), InGaAs (Thorlabs. mod. DET20C)
and LN$_{2}$-cooled InAs (Hamamatsu mod. P7163) photodiodes for the 200-1000\,nm, 900-1700\,nm, 1000-3100\,nm ranges, respectively.  
In the same bands, the spectra are acquired by using Silicon (Oceanoptics, mod. Redtide650) and InGaAs (Oceanoptics, mod. NIR512) CCD based
spectrometers, and also by using a FT-IR interferometer (Bruker, mod. Equinox 55) coupled with the InAs photodiode.

As we have shown in \cite{TmNd}, the LY in the optical range $[\lambda_i,\lambda_f]$,
that is defined by the photodiode responsivity $\eta(\lambda)$ and by the transmittance $T(\lambda) $ of the optical filters used, can be calculated as
\begin{eqnarray}
\mathrm{LY} [\lambda_i,\lambda_f] &=&   \frac{1}{k} \frac{4 \pi d^2}{S} \frac{Q_{d}}{Q_{bs}} \times\nonumber \\
& & \times\left[\left(\int_{\lambda_i}^{\lambda_f} I(\lambda) \lambda d\lambda \right) \bigg/ \left( \int \eta(\lambda) I(\lambda)T(\lambda)
\lambda d\lambda \right) \right]
\label{eqn:LY}
\end{eqnarray} 
Here, $Q_d$ is the total integrated detector response to the electron gun pulse charge $Q_{bs}$,
 $d$ the distance between the crystal and the detector of area $S$ and $I(\lambda)$ the power spectral density of the scintillation.

The normalization factor $k$ only depends on the type of excitation employed, either electrons of X-rays, but does not depend on what detector or crystal are used. 
The IR spectrum can only be investigated by cathodoluminescence because it is much more intense than radioluminescence in our experimental setup.
\eqnref{eqn:LY} thus provides the relative value of the integrated light per unit injected charge in the different bands of the three photodiodes and allows us to merge the spectra, thereby obtaining the scintillation spectral density in the extended wavelength band between $200\,$nm and $3100\,$nm.
 
In order to determine the absolute visible LY for X-rays excitation, the normalisation factor $k$ is estimated by measuring the $Q_d/Q_{bs}$ ratio of a calibrated Pr:LuYAG (0.16\%) scintillator~\cite{Pr,Pr27000}.  The LY in the whole wavelength range is then obtained by exploiting the features of the cathodoluminescence spectra.

\section{Experimental results and discussion}

Our results are presented in two different subsections on the basis of the excitation type. We report the response of the crystals 
to the electron impact excitation and the cathodoluminescence spectra in the first subsection, whereas the value of the LY in the different bands in the second one. 

\subsection{Cathodoluminescence}

\begin{figure}[b!]
\centering
\includegraphics[width=0.50\columnwidth, angle=-90]{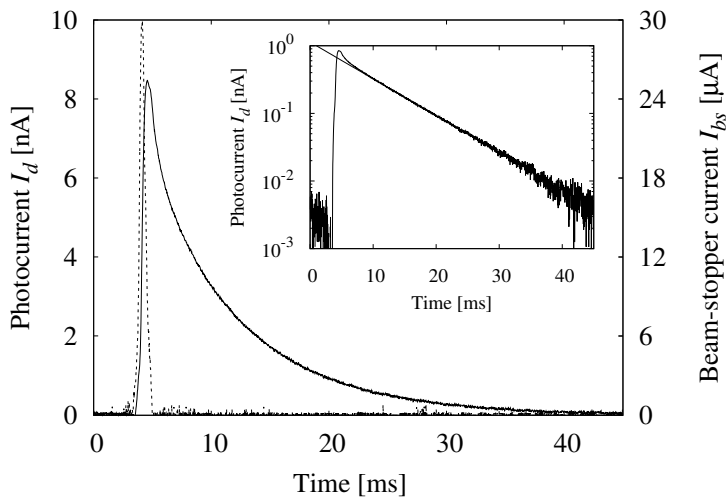}
\caption{\small Time profile of the electron pulse (dashed line, right scale) and the relative time evolution of Ho:YAG cathodoluminescence (solid line, left scale) in the mid-infrared range acquired with the
InAs photodiode. In the inset $I_{d}$ response is shown in semilogarithmic scale in order to put into evidence its monoexponential decay.\label{fig:forma}}
\end{figure}

\begin{figure}[b!]
\centering
\includegraphics[width=0.50\columnwidth, angle=-90]{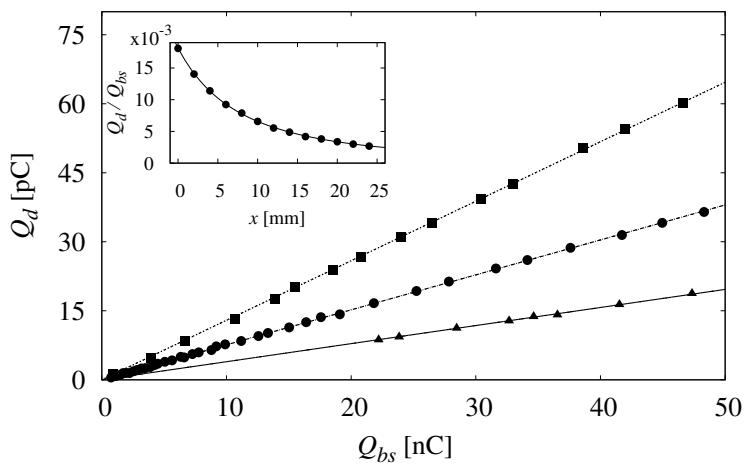}
\caption{\small Silicon (squares), InGaAs (circles) and InAs (triangles) detector response $Q_d$ to the Ho:YLF cathodoluminescence
induced by charge pulse $Q_{bs}$. In the inset, the dependence of the $Q_d/Q_{bs}$ ratio on the crystal-detector distance is shown. The solid line is a fit to \eqnref{eqn:distanza}. The error bars are of the same size of the symbols. \label{fig:lin}}
\end{figure}

Our apparatus allows us to record also the time evolution and the intensity of both electron current injection $I_{bs}(t)$ and  detector 
photocurrent $I_d(t)$. The bandwidth of the transimpedance amplifier (Femto, mod. DLPCA-200) and the characteristic time of the electron pulse allow us to accurately estimate the crystals emission lifetime 
 when it is longer than a few tens of microseconds. We show in \figref{fig:forma}
typical $I_{bs}$ and $I_{d}$ signals. 
Their offline time integration provides the values of $Q_{bs}$ and $Q_{d}$. 
As shown in \figref{fig:lin}, there is a linear relationship between them for all the wavelength bands investigated in the Ho:YLF crystal.

\begin{figure}[b!]
\centering
\includegraphics[width=0.50\columnwidth, angle=-90]{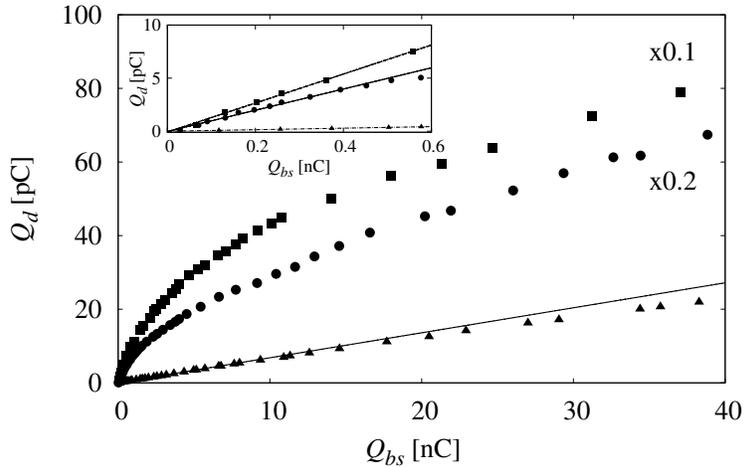}
\caption{\small Response of the Ho:YAG cathodoluminescence to the charge injection the the following bands: $520-580\,$nm (triangles), $670-730\,$nm (circles), $1500-3000\,$nm (squares). Inset: linearity of the response at low excitation level.
 The error bars are of the same size of the symbols. \label{fig:lin2}}
\end{figure}

On the contrary, the scintillation response of the Ho:YAG is linear only for small amplitude of the charge pulse $Q_{bs}$ and 
shows a saturation effect for larger charge injection as one can see by inspecting  \figref{fig:lin2}.  
In this latter case, the following analysis is restricted to the low excitation range and possible explanations of 
the saturation phenomenon will be considered in the Appendix.

The ratio $Q_d/Q_{bs}$, obtained by a linear fit, is also measured as a function of the relative distance $x$ of the detector 
from the crystal in order to obtain an accurate value of the solid angle subtended by different photodiodes. 
Actually, assuming the point source approximation, the data are fitted to the following formula
\begin{equation}
\frac{Q_d(x)}{Q_{bs}} = \left( \frac{Q_d d^2}{Q_{bs}}  \right) \frac{1}{(x+x_0)^2} = \frac{a}{(x+x_0)^2}
\label{eqn:distanza}
\end{equation}
thereby estimating the $a = (Q_d/Q_{bs})  d^2$  
parameter with an accuracy within 5\% for all set of data (see, for instance, the inset in \figref{fig:lin}).

\begin{figure}[b!]
\centering
\includegraphics[width=0.50\columnwidth, angle=-90]{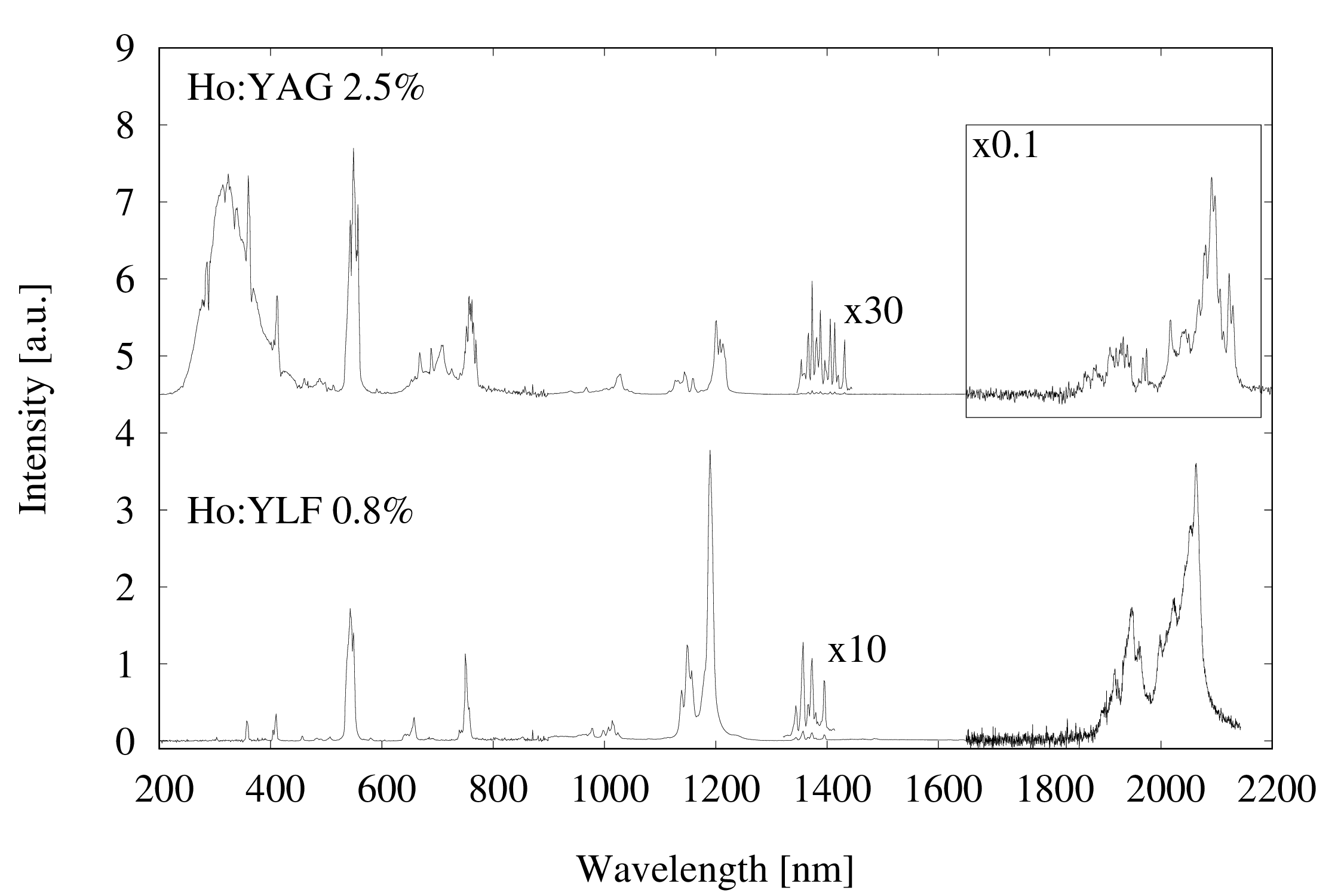}
\caption{\small Cathodoluminescence spectra of Ho:YLF (bottom) and Ho:YAG (top, shifted for clarity) crystals. The electron gun in continuous mode is used
 to improve the signal-to-noise ratio. In the case of Ho:YAG, low electron current intensities ensure that the saturation effects
 do not influence the spectrum. \label{fig:spettro}}
\end{figure}

The parameter $(Q_d/Q_{bs}) d^2$ measured by using Silicon, InGaAs and InAs photodiodes coupled with appropriate longpass and 
shortpass filter, allow us to normalize  the spectra acquired in different wavelength range by means of \eqnref{eqn:LY}.
The resulting emission spectra of Ho:YAG and Ho:YLF are displayed in \figref{fig:spettro}.
We only show the spectral range up to to $2200\,$nm because no emission above the thermal background has been observed.

\subsubsection{Spectroscopic analysis}
\begin{figure}[b!]
\centering
\includegraphics[width=0.50\columnwidth]{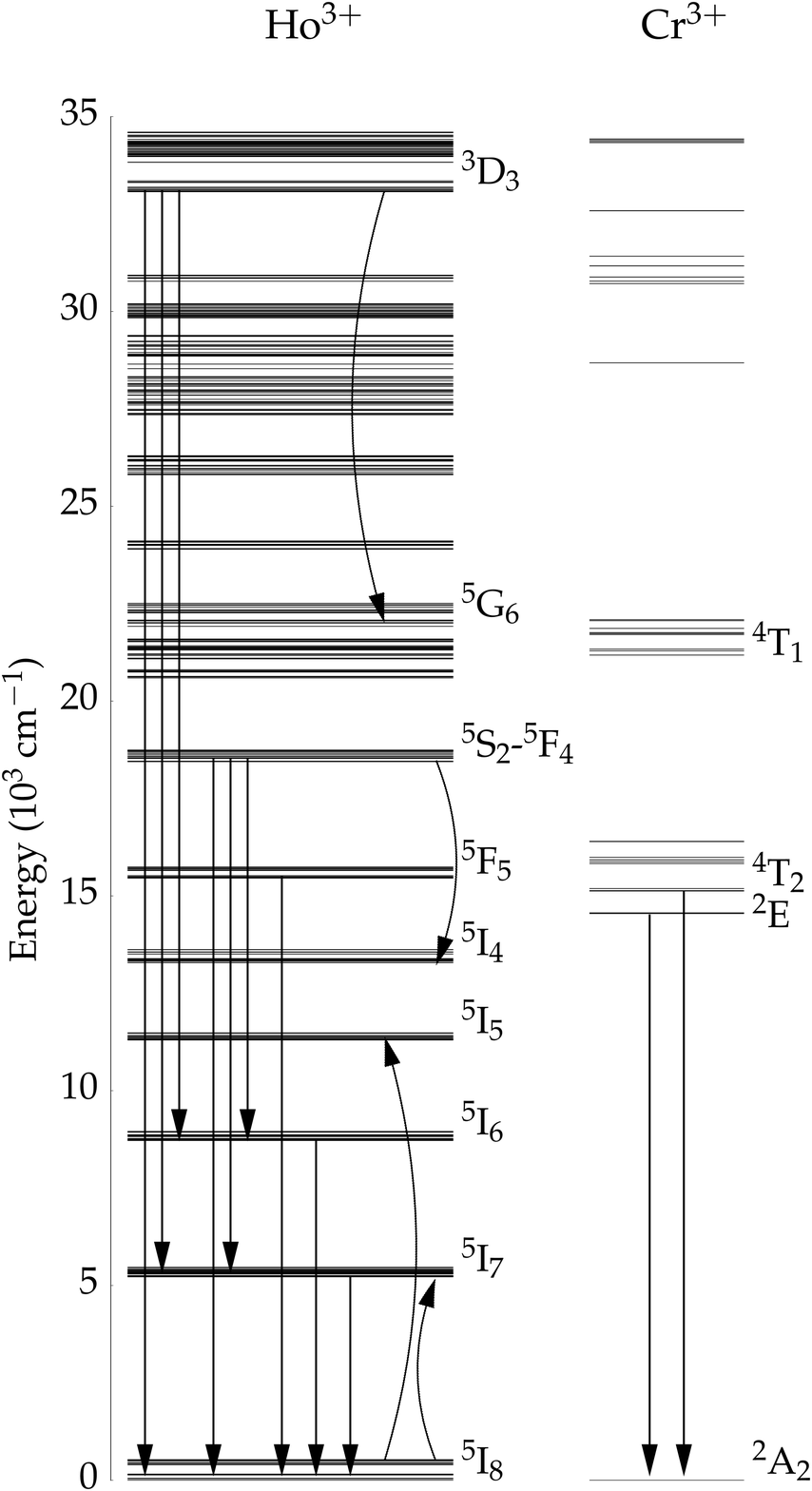}
\caption{\small Scheme of the Ho$^{3+}$~\cite{HoYAGlevel} and Cr$^{3+}$~\cite{Cr_level} calculated levels in the YAG matrix. Only the strongest radiative transitions are shown with straight arrows. The cross relaxation pathways are indicated with curve arrows.\label{fig:livelli}}
\end{figure}

The spectra of the two cystals present several features at the same wavelengths which favorably compare with laser induced fluorescence from similar
crystals both in visible \cite{highlevel, HoYAGVIS, HoYLFVIS} and infrared range \cite{HoYLFINF}. These emissions are attributed to the radiative relaxation of Holmium $4f$ manifolds and
their identification can be found in \tabref{tab:transition} and diplayed in \figref{fig:livelli}.

The Ho:YAG spectrum is characterized by two more emissions than Ho:\-YLF. The first one is located at about $700\,$nm is typical and is widely attributed to 
Cr$^{3+}$ ions in YAG \cite{Cr_700}. The broadband feature in 250-$400\,$nm range
is the same observed by exciting with X-rays undoped YAG\,\cite{YAGemission} and lightly Ho doped oxide crystals\,\cite{HoLuAG} and it is due 
to the matrix exciton emission. 

The lack of self trapped exciton emission in the fluoride crystal, expected in the 200-300\,nm range~\cite{erbioVV}, is probably caused by a more
 efficient energy transfer from excitons to Holmium ions than in the Ho:YAG crystal.
This energy transfer process and the consequent multiphonon relaxation lead to the population of the  $^3\mathrm{D}_3$ manifold, whose emission is heavily quenched by cross relaxation~\cite{highlevel}, in particular by the resonant process
($^3\mathrm{D}_3$,$^5\mathrm{I}_8$)$\rightarrow$($^5\mathrm{I}_5$,$^5\mathrm{G}_6$). It is expected that most of the ions in $^5\mathrm{G}_6$ 
states non radiatively relax to the $^5\mathrm{S}_2$ and $^5\mathrm{F}_4$ states where another efficient cross relaxation process ($^5\mathrm{S}_2$,$^5\mathrm{I}_8$)$\rightarrow$($^5\mathrm{I}_4$,$^5\mathrm{I}_7$) occurs. In the case of Ho:YAG, 
energy transfer between Ho$^{3+}$ in 4f manifolds and Cr$^{3+}$ ions are also possible as it is reported for Ho:Cr:YAlO$_3$ crystals~\cite{Cr_transfer}. 

The multiphonon relaxation dominates for $^5\mathrm{I}_4$ and $^5\mathrm{I}_5$ manifolds, thereby increasing the population of
$^5\mathrm{I}_6$. On the contrary, for the $^5\mathrm{I}_6$ manifold the radiative emission represents a competitive process reducing 
the multiphonon relaxation towards the $^5\mathrm{I}_7$ manifold and the following emission in the mid-infrared range.
In accordance with the fact that the multiphonon relaxation rate is much lower in fluorides than in oxides matrices, 
we measure a lifetime for $^5\mathrm{I}_6$ manifold of ($44 \pm 3$)\,$\mu$s in YAG and ($1.8 \pm 0.1$)\,ms in YLF crystals. Similar
values can be found in literature \cite{YLFrateI6}. This consideration rationalizes the observation that the ratio of the total intensity irradiated in the $1100-1200\,$nm band, stemming from the $^5\mathrm{I}_6 \rightarrow {}^5\mathrm{I}_8$ transition, to that in the $1800-2200\,$nm band, resulting from the $^5\mathrm{I}_7 \rightarrow {}^5\mathrm{I}_8$ transition,  is very different for the two investigated crystals.

The occurence of cross relaxation, that is able to excite up to 3 Holmium ions in low energy manifolds $^5\mathrm{I}_6$ and $^5\mathrm{I}_7$ 
from each ions in high energy $^3\mathrm{D}_3$
can explain the stronger emission in infrared band than in the visible and ultraviolet ones.

After the higher energy manifolds have relaxed, the Ho$^{3+}$ manifold $^5\mathrm{I}_7$ relaxes to the ground state with a single exponential lifetime of
($7.8\, \pm\, 0.1$)\,ms and ($13.5\, \pm\, 0.2$)\,ms in YAG (inset in \figref{fig:forma}) and YLF matrix, 
respectively, in quite good agreement with the values reported in literature~\cite{lifeHoYAG,lifeHoYLF}.

It should be noted that the Holmium ions 5d-4f transitions, expected in the  $160-200\,$nm range for fluoride crystals \cite{HoYLFUV}, 
cannot be observed with our apparatus.

\begin{table}[t!]
\centering
\caption{The involved manifolds and wavelength emission range of the observed Ho$^{3+}$ transitions.}
\begin{tabular}{cc|cc}
\toprule
Transition & emission [nm] & Transition & emission [nm] \\
\midrule
$^3\mathrm{D}_{3}$ $\rightarrow$  $^5\mathrm{I}_{8}$	& 300	&  $^5\mathrm{S}_{2}$-$^5\mathrm{F}_{4}$ $\rightarrow$  $^5\mathrm{I}_{6}$	&	1000-1050 \\
$^3\mathrm{D}_{3}$ $\rightarrow$  $^5\mathrm{I}_{7}$	&	360 & 	$^5\mathrm{S}_{2}$-$^5\mathrm{F}_{4}$  $\rightarrow$ $^5\mathrm{I}_{5}$	&	1340-1450\\
$^3\mathrm{D}_{3}$ $\rightarrow$   $^5\mathrm{I}_{6}$	&	410 & $^5\mathrm{F}_{5}$ $\rightarrow$  $^5\mathrm{I}_{8}$	&	640-660 \\
$^3\mathrm{D}_{3}$ $\rightarrow$   $^5\mathrm{I}_{5}$	&	460  & $^5\mathrm{F}_{5}$ $\rightarrow$  $^5\mathrm{I}_{7}$	&	950-1000   \\
$^5\mathrm{S}_{2}$-$^5\mathrm{F}_{4}$ $\rightarrow$  $^5\mathrm{I}_{8}$	&	530-550 &  $^5\mathrm{I}_{6}$  $\rightarrow$  $^5\mathrm{I}_{8}$	&1130-1210\\
$^5\mathrm{S}_{2}$-$^5\mathrm{F}_{4}$  $\rightarrow$ $^5\mathrm{I}_{7}$	&	740-760 & $^5\mathrm{I}_{7}$   $\rightarrow$ $^5\mathrm{I}_{8}$	& 1800-2100\\

\bottomrule
\end{tabular}
\label{tab:transition}
\end{table}

\subsection{X-ray light yield}
We assume that the emission spectra obtained with electron impact or 
X-ray excitation are the same. The validity of this hypothesis is supported by the fact that X-rays produce photoelectrons 
in the crystals and it has previously been verified for Nd:YAG crystal \cite{TmNd}.

The LY determination is accomplished by following the procedure 
outlined elsewhere~\cite{TmNd}. The measure of the ratio $Q_d/Q_{bs}$ is repeated exciting with X-rays the two Holmium
doped crystals and a calibrated Pr:LuYAG (0.16\%) scintillator whose LY we reported to be $33 $\,photons/keV in the 200-1100\,nm 
band\cite{Pr,Pr27000}. 
An active integrator with a time constant of $\approx 480 \,\mu$s and conversion factor $0.25\,$mV/fC is used in order to improve the signal to noise ratio. 
The linearity between excitation charge and scintillation is observed even for Ho:YAG crystal because of the low excitation density.

 By comparing the luminescence intensity with that of the calibrated crystal through
 \eqnref{eqn:LY}, we can estimate the visible LY of the Ho doped crystals. Then, the infrared LY is inferred from the cathodoluminescence spectra.
 Reabsorption is estimated to be of a few \% at most and is, thus, neglected.

The values obtained for different bands have an overall accuracy better than 15\% and they are shown in \tabref{tab:LY}. Both crystals have low LY in UV-visible range, less than $2\,$photons/keV, whereas almost $40\,$photons/keV are 
emitted in the infrared range. Taking into account the calculated branching
 ratio for the $^5\mathrm{I}_{6}$ manifold \cite{lifeHoYLF,YLFrateI6}, we can infer a LY at 2800\,nm due to the 
 $^5\mathrm{I}_{6}\rightarrow {}^5\mathrm{I}_{7}$ transition of 0.5\,photons/keV for Ho:YLF and a negligible value for Ho:YAG.

Fujimoto {\em et al.} reported $(14.3\,\pm\,1.4)\,$photons/keV for undoped YAG single crystals. As we measure $(1.0\,\pm\,0.2)\,$photons/keV 
for the same band in Ho:YAG 2.5\%, we can conclude that the matrix exciton-Holmium and Chromium ions energy transfer occurs with high
 efficiency. As the energy transfer and cross relaxation processes dominate in these crystals, we do not expect much higher infrared LY in crystals more heavily doped with Ho$^{3+}$, in which concentration quenching could also occur.
 
\begin{table}[t!]
\centering
\caption{Ho:YAG and Ho:YLF LY in the different wavelength ranges. }
\begin{tabular}{ccc}
\toprule
 & \multicolumn{2}{c}{LY [photons/keV]} \\
\midrule
 Wavelength range [nm] & \textbf{Ho:YLF} & \textbf{Ho:YAG} \\
\midrule
200-500 & 0.1 & 1.0 \\
500-800 & 1.4 & 0.7 \\
1100-1300 & 5.4 & 0.3 \\
1800-2200 & 32.8 & 38.8\\
\bottomrule
\end{tabular}
\label{tab:LY}
\end{table}

\section{Conclusions}
The investigation of the feasibility of a rare event detector in which lasers can greatly increase the LY of the active medium,
 has led us to study the infrared scintillation of rare earth doped crystals. In this paper, we have presented the cathodoluminescence spectra and 
 we have estimated the LY of Ho:YAG (2.5\%) and Ho:YLF (0.8\%) crystals in a broad band between 200\,nm and 2300\,nm. 

As the quantum efficiency of the lowest excited manifold $^5\mathrm{I}_7$ is almost 1, the reported LYs in $1800-2200\,$nm range directly
 provide the number of Ho$^{3+}$ ions per keV that are excited into this manifold. 
The obtained high values, together with those related to Tm:YAG and Nd:YAG crystals \cite{TmNd}, build a strong case that particle excitation can 
efficiently populate the low energy, metastable rare earth ion manifolds.

Surely, cross relaxation is an efficient process in which a high lying excited level populates the low lying ones but we cannot claim
if other mechanisms take place. Secondary electrons, with not enough energy to create electron-hole pair, could still excite the RE ions
into their low lying levels. In addition, it is well known that less than half of the particle energy can be converted into visible radiation~\cite{dorenbosintro},
the remaining part creating a large number of phonons that could equally populate the low energy RE ion levels.

At the best of our knowledge there are no studies about these topics. We believe that the investigation of the infrared light yield
could be a useful mean to study the radiation-matter interaction.

For these reasons, and also for the realization of a suitable detector, in the next future, we plan to investigate the temperature dependence o of the LY in several wavelength ranges and, therefore, of the number of ions excited in the different manifolds for crystals as a function of the dopant concentration.

\appendix
\section{Saturation behaviour}

Chromium impurities are common in aluminate crystals and we believe that their presence is responsible for the nonlinearity of Ho:YAG
cathodoluminescence. A similar effect has been reported for the emission of Cr in ZnGa$_2$O$_4$~\cite{rach2001}.

Actually, a high population inversion of the Ho$^{3+}$ activators can be ruled out as a cause of the saturation effect because in the present experiment we used a low current density and because no saturation shows up in the Ho:YLF crystal which has a lower dopant concentration but displays a similar luminescence efficiency.  

We have observed that the intensity of the UV exciton emission and of the Ho$^{3+}$ $^5\mathrm{S}_2$ manifold luminescence share the same dependence on the injected electron  charge. Of the many processes that cause the excitonic decay, there must be at least one that becomes more efficient as the electron gun excitation increases, thereby making less likely other competitive mechanisms such as exciton radiative recombination and exciton-Ho$^{3+}$ ions energy transfer.

One of these excitation intensity dependent processes might be the ionization Cr$^{3+}\!\rightarrow\,$Cr$^{2+}$. Actually, Cr$^{3+}$ ion is an efficient electron trap~\cite{phoph1}. As the charge transfer O$^{2-}\rightarrow$Cr$^{3+}$ band lies at 230\,nm \cite{wang2012}, just above the self trapped exciton emission, it is therefore reasonable that ionization can occur more likely for Cr$^{3+}$ in a excited state than in ground state.
As the current injection increases, more Cr$^{3+}$ ions are excited from the ground state $^4\mathrm{A}_2$ into the metastable manifold
$^2\mathrm{E}$.  The host excitons more likely give up their energy in the O$^{2-}\rightarrow$Cr$^{3+}$ charge transfer rather than in the direct excitation Ho$^{3+}$ ions or in the radiative decay.

We have also observed saturation in the $680-720\,$nm, $1100-1200\,$nm, and $1800-2200\,$nm bands. The charge transfer process probably explains the saturation behaviour of the exciton- and  Ho$^{3+}$ $^5\mathrm{S}_2$ emission, as well as the emission in the $680-720\,$nm range. Actually, some Cr$^{3+}$ ions in the metastable state $^2\mathrm{E}$ are ionized to the Cr$^{2+}$ state instead of relaxing and emitting in near infrared. These ions quickly and non radiatively lose an electron but do not necessary return to $^2\mathrm{E}$ excited levels, thereby decreasing the luminescent efficiency. 

Cr$^{3+}$ ions in YAG present broad band absorptions at $600\,$nm and at $450\,$nm.
We believe that Cr$^{3+}$ in the $^2\mathrm{E}$ manifold with a long lifetime could give also rise to a depopulation of low lying Ho$^{3+}$ manifolds via the energy transfer upconversion (Ho$^{3+}$ $^5\mathrm{I}_{7,6}$, Cr$^{3+}$ $^2\mathrm{E}$)$\rightarrow$(Ho$^{3+}$ $^5\mathrm{I}_{8}$, Cr$^{3+}$ $^4\mathrm{T}_1$), thereby quenching the Ho$^{3+}$ infrared scintillation. Actually, we have characterized the $^{2}\mathrm{E}$ emission by its lifetime of $1.05\,\pm 0.03\,$ms, in quite good agreement with literature data for Cr:YAG crystals at room temperature~\cite{Cr_16}.

These hypotheses can explain why a similar saturation behaviour is also shown by Er:YAG 0.5\%, whose cathodoluminescence displays the same Cr$^{3+}$ emission at 700\,nm \cite{Borghesani2015}, whereas the Tm:YAG and Nd:YAG response in the same excitation condition is linear~\cite{TmNd}. Actually, in the latter cases Cr$^{3+}$ impurities are equally expected to be present but the $^2\mathrm{E}$ emission is missing.  The lifetime of $^{2}\mathrm{E}$ state is strongly shorten because efficient energy transfers occur in Tm$^{3+}$ or Nd$^{3+}$ doped crystals.

\section*{Acknowlegments}
This work is supported by Istituto Nazionale di Fisica Nucleare (INFN) within AXIOMA project. We thank C. Braggio and M. Guarise of the
Padova and Ferrara University respectively for the useful discussion, Prof. M. Tonelli and Prof. A. Di Lieto
of the Pisa University for providing us with the Ho:YLF crystal.

\end{document}